\PassOptionsToPackage{table,dvipsnames}{xcolor}
\documentclass[letterpaper,twocolumn,10pt]{article}
\usepackage{usenix-2020-09}
\pdfoutput=1

\usepackage{colortbl}
\usepackage{amsmath,amssymb,amsfonts}
\usepackage{xspace}
\usepackage{multirow}
\usepackage{soul} % highlighting
\usepackage{color}
\usepackage[newfloat,frozencache]{minted}
\newcommand{\lfencejmp}{\textsc{Lfence/Jmp}\xspace}
\newcommand{\lfence}{\textsc{Lfence}\xspace}
\newcommand{\flushreload}{\textsc{Flush+Reload}\xspace}

\usepackage{pifont}
\newcommand{\bwcmark}{\ding{51}}
\newcommand{\bwxmark}{\ding{55}}
\newcommand{\cmark}{\color{ForestGreen}\bwcmark}
\newcommand{\xmark}{\color{red}\bwxmark}

\usepackage{graphicx}
\microtypecontext{spacing=nonfrench}

\begin{document}
%-------------------------------------------------------------------------------

%don't want date printed
\date{}

\title{You Cannot Always Win the Race:\\Analyzing the \lfencejmp Mitigation for Branch Target Injection}

\author{
{\rm Alyssa Milburn}\\
Intel\footnotemark[1]
\and
{\rm Ke Sun}\\
Intel\footnotemark[1]
\and
{\rm Henrique Kawakami}\\
Intel\footnotemark[1]
}

\maketitle

\footnotetext[1]{This preprint describes work by individual researchers within Intel STORM. It is not intended as documentation or guidance.}

%-------------------------------------------------------------------------------
\begin{abstract}
%-------------------------------------------------------------------------------
\lfencejmp is an existing software mitigation option for Branch Target Injection (BTI) and similar transient execution attacks stemming from indirect branch predictions, which is commonly used on AMD processors. However, the effectiveness of this mitigation can be compromised by the inherent race condition between the speculative execution of the predicted target and the architectural resolution of the intended target, since this can create a window in which code can still be transiently executed.
This work investigates the potential sources of latency that may contribute to such a speculation window. We show that an attacker can ``win the race'', and thus that this window can still be sufficient to allow exploitation of BTI-style attacks on a variety of different x86 CPUs, despite the presence of the \lfencejmp mitigation.

\end{abstract}

%-------------------------------------------------------------------------------
\section{Introduction}
%-------------------------------------------------------------------------------

Indirect branch target predictions allow processors to predict the targets of indirect branches before they are resolved, steering speculative execution to locations of predicted branch targets. These predicted targets are typically based on previous execution history. 
Transient execution attacks using the indirect branch predictor -- such as Branch Target Injection (BTI, aka Spectre v2) and later variants -- involve an attacker attempting to manipulate such predictions in a way that can allow them to transiently execute code which can reveal sensitive information across security domains via an incidental channel, such as the cache state.

Such attacks can generally be mitigated using software techniques (such as retpoline\cite{googleretpoline}), as well as hardware mitigations where available. Recent Intel CPUs support a mitigation (enhanced IBRS, or eIBRS, which we discuss below) that isolates branch prediction entries for more privileged domains, preventing userspace attackers from directly injecting targets for the indirect branch predictions in kernel space.

However, recent research on Branch History Injection (BHI) \cite{bhi} showed that an attacker can still use branch history to influence the target predictions for indirect branches in more privileged code. This research shows that variants of BTI-style attacks may still be possible despite the use of eIBRS, in situations where an attacker can find (or create) disclosure gadgets among existing privileged-mode branch prediction targets.
The BHI research sparked renewed interest in alternative software mitigations for BTI-style attacks.

One such alternative is the \lfencejmp software mitigation for x86 CPUs, which was rejected in favor of retpoline (and more recently, eIBRS) on Intel CPUs. However, it has been documented by AMD as an effective retpoline alternative; in fact, the default Linux kernel mitigation on AMD processors (at the time of writing) is \lfencejmp, referred to as the ``AMD retpoline''. Intel's ecosystem partners requested revisiting this mitigation, to evaluate whether it could be a suitable alternative to retpoline on Intel CPUs.

However, this particular mitigation relies on an inherent race condition, and we show that the remaining window for speculative execution can still allow the transient execution of disclosure gadgets. Especially when combined with windowing gadgets in the form of SMT (Simultaneous Multithreading) workloads running on a sibling logical processor, this speculation window can fit many forms of ``universal read'' \cite{spectreisheretostay} type gadgets, on a variety of x86 CPUs from both Intel and AMD.

This means that, where eIBRS is unavailable or BHI attacks are a concern, \lfencejmp may not be sufficient to mitigate BTI-style attacks.

Our contributions are:
\begin{itemize}
\item We analyze the \lfencejmp mitigation and identify several potential sources of latency which could contribute to a speculation window.
\item We show that a race condition allows transient execution of code, and characterize the resulting speculation window, on a range of CPUs.
\item We develop and test appropriate SMT workloads for several different CPUs, and show that the \lfencejmp speculation window can be significantly expanded using such workloads.
\item We discuss the remaining obstacles for practical exploitation, and demonstrate that attacks are practical by implementing a proof-of-concept exploit which can be used to infer kernel memory contents using unprivileged eBPF on an AMD Zen 2 CPU, despite the use of \lfencejmp.
\end{itemize}

AMD has told us they plan to update their mitigation guidance in response to our work, and have submitted Linux kernel patches to switch to retpoline (instead of \lfencejmp) as the default BTI mitigation for AMD processors.

%-------------------------------------------------------------------------------
\section{Background}
%-------------------------------------------------------------------------------

\subsection{\lfence}
\lfence is an x86 instruction for serializing execution, which has become widely used as a method for mitigating Spectre Variant 1 vulnerabilities in software. Both Intel and AMD documentation describes that instructions after \lfence will not be executed until the instructions before \lfence have completed execution, and the results of those instructions (such as memory loads) are available. We have not observed any behavior which would be incompatible with this definition.

Intel's Software Developer Manual \cite{sdm} states that ``\lfence does not execute until all prior instructions have completed locally, and no later instruction begins execution until \lfence completes''. Although the AMD64 Architecture Programmer's Manual \cite{apm} states that \lfence only ``assures that the system completes all previous loads before executing subsequent loads'', AMD's
software guidance \cite{amdswguidance} documents
an MSR
which makes
\lfence dispatch-serializing: ``upon encountering an \lfence when the MSR bit is set, dispatch will stop until the \lfence instruction becomes the oldest instruction in the machine''. AMD also documents a CPUID enumeration
for processors where
\lfence is dispatch-serializing by default.

We confirmed that the documented MSR bit is set by default on Linux on the relevant AMD CPUs, and that it was set during our experiments.

\subsection{Branch Target Injection and its variants}
Speculative execution due to indirect branch target predictions forms the basis for several recent attacks, including the well-known Spectre Variant 2 attack \cite{bti} (aka BTI) and SpectreRSB \cite{spectrersb} as well as more recent attacks such as BHI. Transiently executed code (on the speculative path) can leave persistent effects (such as cache state changes) that can be measurable by subsequent architecturally executed code, speculative execution is inherently not necessarily restricted by architectural checks and conditions, and attackers may be able to influence branch predictions across security domains.

Combined, these factors may allow an attacker to mount BTI-style attacks that specify (or influence) the branch prediction of a victim branch and, typically, may allow data to be inferred from the victim security domain by speculatively executing a disclosure gadget and leveraging the cache-timing side channel. 

\subsection{Primitives of BTI-style attacks}
A successful BTI-style attack requires several attack primitives. We summarize the terminology \cite{microsoftmitigating} below, which we will use to discuss the expected mechanism of the \lfencejmp mitigation and its limitations for mitigating BTI-style attacks, including variants such as BHI. 

\footnotetext[2]{Since actual `gadgets' are not required, we also refer to these concepts more generally as speculation and windowing \emph{primitives}.}

\textbf{Speculation Gadget\footnotemark[2]}: an indirect branch in the ``victim's'' security domain, with a branch target prediction which can be specified by the attacker. The original BTI attacks exploited the lack of isolation to directly ``inject'' indirect predictor targets, while BHI attacks instead rely on influencing target selection based on branch history. This indirect branch also needs to be invoked by the attacker, with sufficient attacker-controlled context.

\textbf{Windowing Gadget\footnotemark[2]}: an operation to delay the architectural execution of the victim indirect branch (typically by delaying the resolution of the branch target), thus opening the window for speculative execution at attacker-controlled location. The most common approach involves evicting the memory containing the branch target address from the cache.
%In practice, an actual ``gadget'' is not necessarily a requirement.

\textbf{Disclosure Gadget}: a group of instructions that accesses data and conveys it via a side channel. For example, this is typically code which performs a data-dependent cache load, converting data read by transient speculative execution into persistent and measurable cache state changes.

In a typical ``universal read'' disclosure gadget, which reads memory contents, the attacker also needs to control the memory address being read. This allows such a gadget to load potentially-sensitive data from an attacker-controlled location, and then infer the data (e.g., a subsequent load providing a cache side-channel).

\subsection{Mitigations}
Currently there are both software-based and hardware-based mitigation options which can help mitigate attacks such as BTI. In the x86 ecosystem, both Intel and AMD have provided hardware-based mitigation options such as Indirect Branch Restricted Speculation (IBRS), Single Thread Indirect Branch Predictors (STIBP), and Indirect Branch Predictor Barrier (IBPB) to control or configure the branch prediction behavior of the processor \cite{intelmitigations,amdmitigations}.
Recent Intel CPUs also support enhanced IBRS (eIBRS) which provides hardware support for isolating the indirect branch predictions of more privileged domains, preventing attackers from injecting arbitrary branch predictor targets into such domains.

As an alternative to IBRS, Google proposed the ``retpoline'' mitigation \cite{googleretpoline}, which mitigates indirect branch prediction attacks in software by replacing indirect JMP and CALL instructions with a software sequence. Specifically, these branches are replaced with a return instruction which is forced to be mispredicted to a ``safe'' target. Retpoline is widely used, and although there are caveats on some processors where alternate predictors can be used for RET instructions (such as some Intel Skylake-generation processors \cite{retpoline}), it is believed to be an effective mitigation in most circumstances.

Other software mitigations have also been proposed, such as ``randpoline''~\cite{randpoline}, a non-deterministic software mitigation, as well as
\lfencejmp.

\section{\lfencejmp}
\label{sec:lfencejmp}

An alternative approach for mitigating BTI-style attacks is the ``\lfencejmp'' software mitigation. This refers to a code sequence where an \lfence instruction is used to serialize execution before an indirect branch with a register operand (not a memory address), and thus can be executed without any memory-access latency.
We would expect such a sequence to significantly reduce the potential window for transient execution at the predicted target of the indirect branch, since the architectural target will always be available when the branch instruction is allocated for execution, due to the serialization provided by \lfence.

In practice, this mitigation was not recommended by Intel, presumably due to
the lack of architectural guarantees and the availability of alternative mitigations. Google's motivation for retpoline
was also explicit about this topic; they stated that serialization was insufficient, since ``the speculative execution here is a property of the hardware itself''~\cite{googleretpoline}.

AMD's guidance~\cite{amdswguidance}, on the other hand,
states that \lfencejmp (``mitigation V2-2'') is a suitable mitigation for BTI attacks on ``all AMD processors'' since ``\emph{the speculative execution window is not large enough to be exploited}''.

Despite concern from the Linux community given that \lfencejmp was ``not *quite* good enough'' \cite{mailinglistlfencejmp} on Intel CPUs, AMD confirmed that the mitigation is sufficient \cite{mailinglistlfencejmp}.
At the time of writing, \lfencejmp is the default BTI mitigation for both the Linux kernel and widely-used hypervisors (e.g., KVM and Xen) on AMD processors.

\lfencejmp presents as an attractive software mitigation option since it may have lower performance impact in some situations, and is simpler to implement in software compared to alternatives such as retpoline. These factors prompted our research into the potential exploitability of the speculative window inherent to the \lfencejmp mitigation, as part of evaluating it as an mitigation option for BTI-style attacks (such as BHI) on Intel platforms.

\subsection{Proposed mitigation}
\label{sec:lfencejmpimpl}
While other mitigation options generally rely on restricting the Speculation Primitive of BTI-style attacks, the \lfencejmp mitigation proposal aims instead to remove the Windowing Primitive. In other words, with the \lfencejmp mitigation, attacker-controlled transient execution can still technically happen, but the expectation is there will not be a sufficiently large speculation window for this to be exploitable.

In AMD's documentation, the \lfencejmp mitigation is described as ``convert an indirect branch into a dispatch serializing instruction sequence''. It uses this example sequence:

\begin{minted}[linenos]{nasm}
jmp *[eax] ; jump to address pointed to by EAX
\end{minted}

And gives an equivalent example using \lfencejmp:

\begin{minted}[linenos]{nasm}
mov eax, [eax] ; load target address
lfence ; dispatch serializing instruction
jmp *eax
\end{minted}

Merely adding an \lfence before the original code sequence would be insufficient, since the load from memory could act as a windowing gadget. In this example a memory-based indirect branch which loads its target from memory is converted to a load and a register-based indirect branch, with an \lfence added in between to make sure ``the load has finished before the branch is dispatched''.

\subsection{The \lfencejmp race}

A key component of BTI-style attacks is the race condition between the speculative execution path and the architectural execution path of the indirect branch. In other words, speculative execution of a potential disclosure gadget at the predicted target must occur before the actual target of the indirect branch is resolved and the branch is architecturally executed.

The branch prediction unit (BPU) is part of the front end of the pipeline and is intended to provide branch predictions as early as possible, before instructions are dispatched for execution –- and potentially even before they are decoded. Therefore, there is always the potential for a speculation window before the indirect branch instruction is actually executed, at which point a misprediction will be signaled and the pipeline will be cleared.

Thus, the success of a speculative execution attack partially relies on the time (latency) it takes from the point the indirect branch is fetched/predicted to the point it gets executed. We identified four major potential causes of such latency:

\begin{enumerate}
\item \textbf{Data-dependency}: Caused by the delayed availability of the indirect branch target.

\item \textbf{Single-thread resource contention}: Latency caused by other instructions in the out-of-order pipeline.

\item \textbf{Baseline execution latency}: The inherent latency needed for allocation and execution of the branch.

\item \textbf{SMT resource contention}: Resource contention from the sibling thread, which we discuss below.
\end{enumerate}

\lfence serializes previous instructions with respect to the indirect branch, which ensures that the indirect branch target is available (removing the data-dependency) and that all previous instructions have executed (removing any single-thread resource contention caused by previous instructions).

However, \lfence does not affect the baseline execution latency nor any potential SMT resource contention, leaving potential gaps in the effectiveness of the \lfencejmp mitigation. In other words, instructions can still be transiently executed in the window after the serialization of \lfence, but before the indirect branch instruction is executed and the misprediction is signaled.

\subsubsection{SMT resource contention}

When SMT is in use, hardware resources are shared between the sibling logical processors on the same core. Decoded instructions from both logical processors are dynamically dispatched and executed on execution ports and may result in resource contention and cause measurable delay in execution of certain instructions.

We need to consider the contention for resources which could result in the branch execution on one logical processor being delayed by activity on the sibling logical processor. We theorized about three broad such categories:

\textbf{General port contention}: Instructions may need to be executed by specific execution ports, which is a known source of contention \cite{portcontention}. Executing instructions that can use the same execution port of indirect branch on one logical processor may delay the execution of the indirect branch on the sibling logical processor, allowing the speculative execution path to continue, and thus expanding the speculation window.

\textbf{Branch-specific contention}: There may also be shared units and/or resources that are specific to branch instructions (or perhaps indirect branch instructions). The allocation and execution of branch instructions may depend on the availability of such resources. This means that branches being executed on one logical processor may delay the execution of the indirect branch on the sibling logical processor, expanding the speculation window.

\textbf{Contention from unknown causes}: There are likely causes of contention which are not included above, which we group into a single category. We tested several different workloads to observe the contention behavior which may be introduced for unexpected reasons.

\section{Speculation window experiments}

To analyze the speculation windows which may arise from the \lfencejmp race condition, we conduct two sets of experiments: first investigating the baseline execution latency, and then investigating SMT port contention. These experiments were run on a variety of recent x86 CPUs from both Intel and AMD; the specific processors used for testing are listed in Table~\ref{tab:processors}.

\begin{table}
\centering
\begingroup
\rowcolors{2}{gray!10}{white}
\begin{tabular}{|l l|}
\rowcolor{gray!25}
\hline
Microarchitecture & Tested processor\\
\hline
Goldmont Plus &	Intel Pentium Silver N5000\\
Tremont & Intel Core i5-L16G7 (Lakefield)\\
Sunny Cove & Intel Core i5-1038NG7 (Ice Lake)\\
Willow Cove & Intel Core i7-1165G7 (Tiger Lake)\\
Golden Cove & Intel Core i9-12900K (Alder Lake)\\
Gracemont & Intel Core i9-12900K (Alder Lake)\\
\hline
Zen & AMD Ryzen 5 2400G\\
Zen+ & AMD Ryzen Threadripper 2990WX\\
Zen 2 & AMD Ryzen 7 4700G\\
Zen 3 & AMD Ryzen 5 5600G\\
\hline
\end{tabular}
\endgroup
\caption{Processors used in our experiments, along with the corresponding microarchitectures.}
\label{tab:processors}
\end{table}

Our experiments use a simple userspace test sequence which trains the indirect branch predictors so that the predicted target of an indirect branch (on line 4, below) goes to the location of a disclosure gadget, and then executes the indirect branch with a different target, expecting it to be mispredicted. This follows AMD's guidance (see Section~\ref{sec:lfencejmpimpl}).

\begin{minted}[linenos]{nasm}
mov rcx, target_ptr
mov rcx, [rcx]
lfence
jmp *rcx
\end{minted}

One possible issue with conducting these experiments in userspace is that OS context switches may happen between the \lfence and the \textsc{Jmp}, and instructions such as SYSRET are not documented to be serializing on AMD processors. In situations where there may be doubt (low success rates), we reproduced the relevant results in a kernel mode environment with interrupts disabled.

All the experiments which did not involve SMT port contention were run on a single logical processor, with the SMT sibling thread disabled, to rule out the possibility of unintended contention. We also designed the code to rule out false positives due to prefetching or other unexpected sources of speculation.
However, bear in mind that it is likely to be possible to improve / optimize
many of our experiments.

\subsection{Observing the race condition}
\label{sec:observingrace}

\begin{table*}[t]
  \centering
  \begingroup
  \begin{tabular}{|l  c c  c c|}
    \hline
    \multirow{2}{8em}{Microarchitecture} & \multicolumn{2}{c}{\emph{Load-shift-load} gadget} & \multicolumn{2}{c|}{\emph{Load-once} gadget}\\
    & No \lfence & With \lfence & No \lfence & With \lfence\\
    \hline
    Goldmont Plus & >99\% & 0\% & >99\% & \textasciitilde10\%\\
    \rowcolor{gray!10}
    Tremont & >99\% & 0\% & >99\% & \textasciitilde50\%\\
    Sunny Cove (ICL) & >99\% & 0\% & >99\% & >99\%\\
    \rowcolor{gray!10}
    Willow Cove (TGL) & >99\% & 0\% & >99\% & >99\%\\
    Golden Cove (ADL) & >99\% & 0\% & >99\% & >99\%\\
    \rowcolor{gray!10}
    Gracemont (ADL) & >99\% & 0\% & >99\% & >99\%\\
    Zen & >99\% & 0\% & >99\% & >99\%\\
    \rowcolor{gray!10}
    Zen+ & >99\% & 0\% & >99\% & >99\%\\
    Zen 2 & >99\% & 0\% & >99\% & >99\%\\
    \rowcolor{gray!10}
    Zen 3 & >99\% & 0\% & >99\% & >99\%\\
    \hline
  \end{tabular}
  \endgroup
  \caption{Verification of the effectiveness of \lfencejmp, and confirmation of the race condition.}
  \label{tab:experiment1}
\end{table*}

\begin{figure*}[t]
  \centering
  \includegraphics[width=0.8\textwidth]{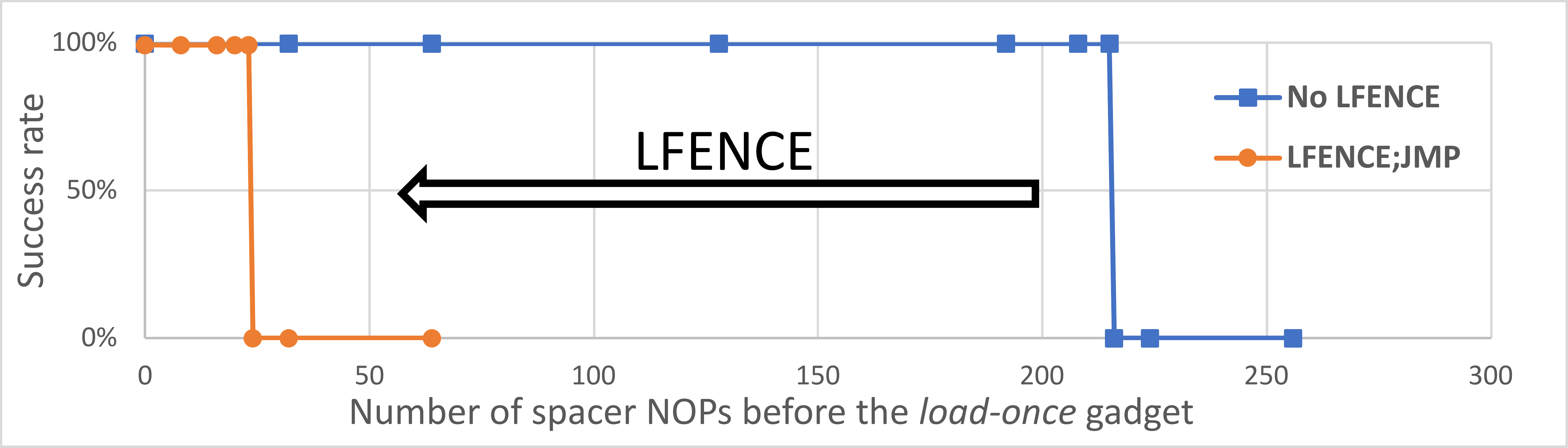}
  \caption{Typical effect of \lfencejmp on baseline execution latency (example on Zen 2).}
  \label{fig:lfencereduction}
\end{figure*}

As discussed in Section~\ref{sec:lfencejmp}, the \lfencejmp mitigation essentially relies on a race condition. There will be a speculative window created by the baseline execution latency, which is the inherent latency needed for the indirect branch execution. We investigated this by attempting to execute some disclosure gadgets within this speculation window on many Intel and AMD CPUs.

We tested two types of disclosure gadgets, both of which load a cache line, allowing data to be inferred using a cache-based incidental channel (\flushreload~\cite{flushreload}):

``\textbf{Memory-disclosure}'' gadget: specifically, a \emph{load-shift-load} gadget with two dependent load instructions and a shift instruction. The first loads a value from a memory location using a pointer in a register (line 1), which is then shifted to achieve a one-to-one correspondence between data value and cache lines (line 2), and the second performs a data-dependent load using the shifted value (line 3), which should be visible in the cache after the speculative execution of the gadget:

\begin{minted}[linenos]{nasm}
mov ebx, [rdi + rdx]
shl ebx, 0xc
mov ecx, [rsi + rbx]
\end{minted}

``\textbf{Register-disclosure}'' gadget: here, a \emph{load-once} gadget with one load instruction, which carries out a speculative cache load based on data which is already present in a register:

\begin{minted}{nasm}
mov ecx, [rsi + rdi]
\end{minted}

We observed similar results on
all the processors we tested, as shown in Table~\ref{tab:experiment1}. For the \emph{load-shift-load} gadget, we see a >99\% success rate without adding \lfence, but a 0\% success rate with the \lfencejmp mitigation, indicating the baseline execution latency cannot open a speculation window large enough to execute the ``memory-disclosure'' gadget with two dependent load instructions. For the \emph{load-once} gadget, we see a >99\% success rate for both the cases with and without \lfence (adding the \lfence decreases the success rate by <1\%), indicating the speculation window is large enough to execute a one-instruction ``register-disclosure'' gadget with just the baseline execution latency, despite the use of \lfencejmp. This confirms our expectation that such a speculation window should be present.

\begin{table*}
  \centering
  \begingroup
\begin{tabular}{|l |l  l l l  l  l | l  l  l  l|}
%\rowcolor{gray!25}
\hline
& \multirow{2}{4em}{Goldmont Plus} & Tremont & \multirow{2}{3em}{Sunny Cove} & \multirow{2}{3em}{Willow Cove} & \multirow{2}{3em}{Golden Cove} & Gracemont & Zen & Zen+ & Zen 2 & Zen 3\\
& & & & & & & & & & \\
\hline
% GLC: 8 or 14?
Max \# of NOPs & 5 & 5 & 2 & 2 & 14 & 23 & 23 & 23 & 23 & 15\\
\hline
\end{tabular}
\endgroup
  \caption{NOP spacer experiment results with the \emph{load-once} ``register-disclosure'' gadget and \lfencejmp (without SMT). }
  \label{tab:experimentnop}
\end{table*}

\subsection{Baseline execution latency experiments}
\label{sec:baselinelatency}

To quantitatively characterize the baseline execution latency, our next experiment added 1-byte NOP instructions to the start of the \emph{load-once} gadget, which act as spacers and delay allocation or execution of the MOV instruction. By adding NOPs until we no longer observed any cache changes when running the test, this approach determines the point where the speculation window is no longer large enough to reach and execute the MOV instruction. Although NOP instructions may overestimate the practical size of the speculation window in terms of the ``number of instructions'' (since they may not actually be allocated or executed), this experiment provides an estimate of the ``upper bound'' on the potential size of the window and we did not observe any differences when replacing the NOP instructions with CBW instructions.

Figure~\ref{fig:lfencereduction} presents a typical example of the reduction in baseline execution latency, measured using this method. This visualizes the impact that \lfencejmp can have on the speculation window of an indirect branch. The number of instructions which can be transiently executed is greatly reduced, but a residual window clearly still remains. Note that the real-world latency without \lfence may be significantly different, and this example reflects only one specific test environment on a single CPU.

Table~\ref{tab:experimentnop} provides these NOP experiment results for those tested processors. Specifically, we measured the maximum number of NOPs which we could add in front of the ``register-disclosure'' gadget before the cache effects were no longer visible due to reaching the limit of the speculation window with the \lfencejmp mitigation.

The speculation window for some processors appears to be considerably larger than implied by the results from Section~\ref{sec:observingrace}, in particular for the AMD processors and some recent Intel processors.

Although we only investigated the baseline execution latency of \lfencejmp, the speculation window may also be affected by other (dynamic) factors, such as the indirect branch predictor latency or the alignment of branches/targets, even in the absence of SMT. We have seen such differences during our experiments, and plan to investigate further; Table~\ref{tab:experimentnop} contains the values we typically observed during our experiments.
The speculation window size may also differ between different variants of a microarchitecture; we report only the numbers we observed on the tested processors from Table~\ref{tab:processors}.

\subsection{Fitting in the window}
\label{sec:fittinginthewindow}

The results from Section~\ref{sec:baselinelatency} show that the speculation window on some processors seems to be significant, even when SMT is disabled or unavailable. However, the results in Section~\ref{sec:observingrace} show that our ``memory-disclosure'' gadget does not appear to fit in this window.

If a mitigation only allows a single load to be executed within the window, without the possibility of other instructions consuming the result of the load, then it may be considered an acceptable mitigation in many circumstances (in the absence of other vulnerabilities).

A detailed analysis of exactly which instructions may fit in such a window, and thus which disclosure gadgets are viable, is outside the current scope of this work. However, we briefly investigated potential options for smaller gadgets, some of which may qualify as so-called ``universal read'' gadgets \cite{spectreisheretostay}.

The \emph{load-shift-load} gadget we tested above is more complicated than strictly necessary. We could instead consider a minimal ``memory load'' gadget with a bare-minimum two-dependent-load sequence (e.g., ``mov rax, [rax]'' and then ``mov rax, [rax]'', without use of complex addressing modes or an explicit shift instruction). This does not appear to be sufficient for a ``universal read'' gadget, but could potentially be used to infer a subset of bits (e.g., when trying to read pointer values).

Other variations on ``memory-disclosure'' gadgets could also change the ``size'' of the gadget and allow it to fit in a smaller speculation window. For example, a different memory access instruction (such as a store or prefetch instruction) could be used instead of the second load; we will refer to this (e.g., ``mov rax, [rax]'' and ``movq [rax], 0'') as a ``store gadget'', a variant using \textsc{Prefetcht0} as a ``prefetch gadget'' and one using \textsc{Clflush} as a ``flush gadget'' (with adjustments to the \flushreload channel).

To obtain one form of a minimal ``universal read'' gadget using only two dependent loads, we could remove the explicit shift, but retain the complex addressing, which allows us to use a base pointer -- and potentially also a shift inside the operand of the second load. This means that the attacker no longer has a one-to-one mapping between data and cache lines, but could still allow an attacker to infer some (higher) bits. If the attacker controls the base register, they can slowly increase it, allowing the remaining bits to be determined. We refer to such a gadget (e.g., loading from [rsi + rbx*8]) as a ``complex load gadget''.

\begin{table*}
  \centering
  \begingroup
  \begin{tabular}{|l |l  l  l l l  l | l  l  l  l|}
  \hline
  & \multirow{2}{4em}{Goldmont Plus} & Tremont & \multirow{2}{3em}{Sunny Cove} & \multirow{2}{3em}{Willow Cove} & \multirow{2}{3em}{Golden Cove} & Gracemont & Zen & Zen+ & Zen 2 & Zen 3\\
  & & & & & & & & & & \\
  \hline
  Minimal load & 0\% & 0\% & 0\% & 0\% & 0\% & >99\% & >99\% & >99\% & >99\% & 0\%\\
  \rowcolor{gray!10}
  Store & 0\% & 0\% & 0\% & 0\% & 0\% & 0\% & >99\% & >99\% & >99\% & 0\%\\
  Prefetch & 0\% & 0\% & 0\% & 0\% & 0\% & >99\% & >99\% & >99\% & >99\% & 0\%\\
  \rowcolor{gray!10}
  Flush & 0\% & 0\% & 0\% & 0\% & 0\% & 0\% & 0\% & 0\% & 0\% & 0\%\\
  Complex load & 0\% & 0\% & 0\% & 0\% & 0\% & >99\% & >99\% & >99\% & >99\% & 0\%\\
  \hline
  \end{tabular}
  \endgroup
  %\vspace{0.1cm}\\
  %{\footnotesize \textdagger We also observed a 0\% rate for similar variants of the ``register-disclosure'' gadget on Tremont.}
  %\vspace{-0.3cm}
  \caption{Test success rate for minimal variants of the ``memory-disclosure'' gadget (without SMT).}
  \label{tab:experimentminimal}
\end{table*}

Table~\ref{tab:experimentminimal} shows results for attempting to transiently execute some of these gadgets despite \lfencejmp. Although we only managed to fit any such gadgets in the speculation window on one Intel processor -- Gracemont -- without enabling SMT, all tested AMD parts prior to Zen 3 also have a speculation window large enough to execute a memory dependency after the load, without SMT. 

These results support AMD's own findings (see the Disclosure section) that two dependent loads may fit in the window on AMD processors -- as well as our observation of a reduced speculation window size on Zen 3 (see Table~\ref{tab:experimentnop}). They also corroborate the much smaller baseline execution latency we observed on most Intel CPUs.

There are many potential directions; we will briefly
summarize two of them.
First, we could replace
\flushreload with a different cache-based side channel.
For example, evicting cache lines from the L1 cache (rather than flushing them from the entire cache hierarchy) could avoid the need for the second load to fetch from uncached memory. The size of the needed speculation window could be further reduced by avoiding evicting lines at all; for example, using cache LRU states \cite{lrustates}, which can provide a cache-based side channel for data which is already present in the cache. 

Second, we may be able to remove the need for a second dependent load entirely by using a non-cache/memory-based side channel. For example, an attacker could train the conditional branch predictor in a disclosure gadget, and infer data using the behavior of an aliased branch \cite{branchscope}. However, our attempts so far to do this within the \lfencejmp window (without SMT) have been unsuccessful.

\subsection{SMT port contention experiments}
\label{sec:smtexperiments}

Workloads on the sibling logical processor can act as windowing gadgets, potentially creating an opportunity for attackers to win the race condition and increase the speculation window such that a disclosure gadget can transiently execute.

In practice, we observed that certain workloads with contention for branch-specific resources seems to be able to cause effective SMT port contention with indirect branches in the test on both Intel and AMD platforms.

Multiple branch-focused workloads were evaluated including the ones consisting of direct JMPs, direct far JMPs, indirect near JMPs and conditional JMPs. For each of the last two categories, we created workloads with both correctly-predicted and mispredicted branches.

A XOR-heavy workload was also tested as an example for general port contention.
Moreover, we also experimented with several different ``random'' workloads looking for unknown sources of contention, and present results for the most ``interesting'' workload we discovered in this category, which consists of repeated calls to the nanosleep function in a loop.

Note that we did not attempt to optimize these workloads in terms of the effective instruction density, nor to synchronize them with the indirect branch execution on the sibling logical processor. In particular, the mispredicted branch workloads may be less effective in terms of branch density in the code. This is because these branches need to be trained to be constantly mispredicted, during which period the workloads may not be causing the desired contention.

\begin{table*}
  \centering
  \begingroup
  \begin{tabular}{|l | c | c | c | c| c | c | c | c | c | c | c | c |}
    %\rowcolor{gray!25}
    \hline
    \multirow{3}{6.5em}{Microarchitecture} & \multirow{3}{3.1em}{\centering No \textsc{Lfence}} & \multirow{3}{3.4em}{\centering No workload} & \multicolumn{8}{c|}{SMT workloads}\\
    \cline{4-11}
    & & & \multirow{2}{2.1em}{Direct JMP} & \multicolumn{2}{c|}{Jcc} & \multicolumn{2}{c|}{Indirect JMP} & \multirow{2}{1.7em}{\centering Far JMP} & \multirow{2}{1.8em}{XOR} & \multirow{2}{2em}{nano sleep}\\
    \cline{5-8}
    %& & & & good pred & mispred & good pred & mispred & & & \\
    & & & & predicted & mispred & predicted & mispred & & & \\
    %& No \lfence & With \lfence & No \lfence & With \lfence\\
    \hline
    Sunny Cove (ICL) & \cellcolor{gray!10}{>99\%} & 0\% & 0\% & \cellcolor{gray!10}{<1\%} & 0\% & \cellcolor{gray!10}{98\%} & 0\% & 0\% & \cellcolor{gray!10}{10\%} & \cellcolor{gray!10}{1\%}\\
    %\rowcolor{gray!10}
    Willow Cove (TGL) & \cellcolor{gray!10}{>99\%} & 0\% & 0\% & \cellcolor{gray!10}{<1\%} & 0\% & \cellcolor{gray!10}{95\%} & 0\% & 0\% & \cellcolor{gray!10}{10\%} & \cellcolor{gray!10}{2\%}\\
    Golden Cove (ADL) & \cellcolor{gray!10}{>99\%} & 0\% & 0\% & 0\% & 0\% & \cellcolor{gray!10}{99\%} & 0\% & 0\% & \cellcolor{gray!10}{13\%} & \cellcolor{gray!10}{2\%}\\
    %\rowcolor{gray!10}
    \hline
    Zen & \cellcolor{gray!10}{>98\%} & 0\% & 0\% & 0\% & \cellcolor{gray!10}{8\%} & 0\% & \cellcolor{gray!10}{5\%} & \cellcolor{gray!10}{15\%} & \cellcolor{gray!10}{30\%} & \cellcolor{gray!10}{6\%}\\
    Zen+ & \cellcolor{gray!10}{>99\%} & 0\% & 0\% & 0\% & \cellcolor{gray!10}{12\%} & 0\% & \cellcolor{gray!10}{5\%} & \cellcolor{gray!10}{23\%} & \cellcolor{gray!10}{43\%} & \cellcolor{gray!10}{8\%}\\
    %\rowcolor{gray!10}
    Zen 2 & \cellcolor{gray!10}{>99\%} & 0\% & 0\% & 0\% & \cellcolor{gray!10}{21\%} & 0\% & \cellcolor{gray!10}{6\%} & \cellcolor{gray!10}{11\%} & \cellcolor{gray!10}{<1\%} & \cellcolor{gray!10}{5\%}\\
    Zen 3 & \cellcolor{gray!10}{>99\%} & 0\% & 0\% & 0\% & 0\% & 0\% & 0\% & 0\% & 0\% & \cellcolor{gray!10}{1\%}\\
    \hline
  \end{tabular}
  \endgroup
  \caption{SMT port contention: \emph{load-shift-load} ``memory-disclosure'' gadget success rates with the \lfencejmp mitigation.}
  \label{tab:experimentsmt}
\end{table*}

Table~\ref{tab:experimentsmt} shows results for all the tested processors which support SMT. We only evaluated a ``memory-disclosure'' gadget, since our earlier experiments (Section~\ref{sec:observingrace}) demonstrated that \lfencejmp generally does not prevent transient execution of the ``register-disclosure'' gadget, even in the absence of SMT port contention. On all processors tested, we observed at least one SMT workload which can induce a large enough speculation window to transiently execute the \emph{load-shift-load} gadget, despite the use of the \lfencejmp mitigation.

On AMD microarchitectures prior to Zen 3, we have observed that mispredicted branches (as well as far JMPs) executed on a sibling logical processor appear to be an effective workload to increase latency in the indirect branch execution, which may indicate contention for resources that are specific to mispredicted branches. Similarly, sufficient contention is also seen with our XOR workload on microarchitectures prior to Zen 2, which is likely due to general port contention.

The best SMT workload for the tested Intel CPUs appears to be correctly-predicted indirect JMPs, which may indicate contention for resources specific to indirect branches. The expanded speculation window caused by the nanosleep workload may have the same root cause, since indirect branches will be executed in the kernel on the sibling logical processor. However, this cannot be the only source of contention, since we also observed hits for some other SMT workloads.

On Zen 3, the majority of these SMT workloads do not appear to increase the speculation window after the \lfencejmp branches on the sibling logical processor, which may indicate fewer shared resources or differences in the way in which resources are allocated. However, since the nanosleep workload still opens a sufficiently large speculation window to execute the \emph{load-shift-load} gadget, there still appears to be at least some form of resource contention between sibling logical processors on Zen 3. Although it would presumably be possible to root-cause the source of the contention and develop a more focused workload, we did not consider this necessary given the scope of this study.
We also have not investigated alternative port contention workloads.

\section{Exploitability}

\subsection{Another Dependent Load}
The above experiments show that, with concurrent SMT workloads, the speculative execution window beyond an \lfencejmp sequence can be significantly larger than the window needed for a ``universal read'' disclosure gadget. We discussed some options for smaller speculation windows above, but
for these larger speculation windows, a key question remains: whether an attacker would need to control register values at the indirect branch. If a speculation window is large enough to permit a third dependent load, such control would not be necessary, which could significantly increase the number of viable disclosure gadgets. This would allow, for example, forms of ``universal read'' gadgets which read the address of the desired data by the gadget from the stack, rather than needing it to be in a register.

We implemented an artificial proof-of-concept attack which infers the contents of a string from memory, byte-by-byte, using \flushreload to observe the resulting cache effects. The gadget used is shown below; it is similar to the ``memory-disclosure'' gadget evaluated above, but the target address is read indirectly from memory rather than taken directly from an attacker-controlled register. We also mask the value with 0xFF, which allows easier inference of memory contents on a per-byte basis (similar results could be obtained in a smaller window by just using one instruction such as MOVZBQ):

\begin{minted}[linenos]{nasm}
  mov rbx, [rbx]
  mov rdx, [rbx]
  and rdx, 0xff
  shl rdx, 0xc
  mov rax, [rdx+rcx]  
\end{minted}

We combined this attack with the ``best-performing'' SMT workload (from Table~\ref{tab:experimentsmt}) for each processor; Table~\ref{tab:experimentpoc} shows the results. As can be seen, even in the Zen 3 case (where we have yet to isolate the effective instructions in the SMT workload), this more generic disclosure gadget can fit in the speculation window despite the use of the \lfencejmp mitigation.

\begin{table*}
  \centering
  \begingroup
  \begin{tabular}{|l |l  l  l l  l  l | l  l  l  l|}
  \hline
   & \multirow{2}{4em}{Goldmont Plus} & Tremont & \multirow{2}{3em}{Sunny Cove} & \multirow{2}{3em}{Willow Cove} & \multirow{2}{3em}{Golden Cove} & Gracemont & Zen & Zen+ & Zen 2 & Zen 3\\
  & & & & & & & & & & \\
  \hline
  \rowcolor{gray!10}
  No SMT & \xmark & \xmark & \xmark & \xmark & \xmark & \xmark & \xmark & \xmark & \xmark & \xmark\\
  SMT workload & N/A & N/A & \cmark & \cmark & \cmark & N/A & \cmark & \cmark & \cmark & \cmark\\
  \hline
  \end{tabular}
  \endgroup
  \caption{Proof-of-concept three-load variant results: does the PoC obtain the secret?}
  \label{tab:experimentpoc}
\end{table*}

\subsection{Remaining Obstacles}
The results above alone do not necessarily mean that the speculation window opened by the \lfencejmp race condition is exploitable by an attacker. The practical exploitability depends on many other factors, even assuming that an attacker has already obtained arbitrary code execution in userspace.

The speculation window size may also have other consequences; for example, page walks may not be possible, which may limit speculative memory accesses to pages which have their address translations in the Translation Lookaside Buffer (TLB). Similarly, it may not be practical to execute some instructions due to conflicting needs for pipeline resources.

On the other hand, a mitigation which allows a speculation window containing multiple dependent loads -- potentially allowing transient execution of a cache-based ``universal read'' gadget --
raises a higher level of concern. We have shown that \lfencejmp permits such a window on many CPUs, which may make it unattractive as an alternative for retpoline.

Exploitation in practice also depends on which potential indirect call sites and disclosure gadgets could be used by an attacker. Assuming a scenario where a userspace attacker attempts to obtain data from kernel mode, we assume Supervisor-Mode Execution Prevention (SMEP) is enabled, which requires the disclosure gadgets to be located in executable kernel memory. If predicted targets are not isolated between modes, such as on older Intel CPUs as well as current AMD CPUs (as far as we can determine), a userspace attacker can inject targets into indirect branch predictor entries, and any bytes in executable kernel code may be a potential gadget. Defenses such as fine-grained Address Space Layout Randomization (ASLR) can be bypassed with the use of cache-timing side channels.

On the other hand, if the userspace attacker cannot specify predicted targets for the kernel indirect branches directly (as may be the case with Intel's eIBRS), and potential locations for transient execution are limited to existing kernel targets, then exploitation also requires identifying a suitable disclosure gadget within this more restricted scope.

\subsubsection{AMD target aliasing}

Although previous work has shown that older (pre-eIBRS) Intel CPUs do not isolate predicted targets between modes, allowing injection of targets, we are unaware of any explicit documentation of this behavior on recent AMD CPUs (although \cite{spectre} stated this was true on Zen). Nevertheless, our experiments show that more recent AMD CPUs also seem to lack mode-based target isolation; the only obstacle is that bit 47 is not set for userspace pages.
In particular, we
 experimented with cross-mode aliasing of short indirect branch predictor targets on AMD processors, which was the approach used by the original Spectre Variant 2 attacks on older Intel CPUs without eIBRS~\cite{spectregpz}.
A simpler approach (as discussed in~\cite{spectre}) is to branch to an illegal target and suppress the fault.

Our cross-mode aliasing experiments
% results
confirmed that a userspace attacker without the ability to execute code at a linear address with bit 47 set can instead toggle other address bits of a userspace branch to cause collisions due to aliasing with kernel branches. The branches contributing to the BHB (Branch History Buffer, following the terminology from \cite{spectregpz}), the indirect branch used to train the predicted target as well as the branch target itself, can thus be executed in userspace, allowing an attacker to control the lower bits of the predicted targets of kernel indirect branch addresses. The specific aliasing behavior (and the number of controllable lower bits of the predicted targets) varies between CPU generations, but we reproduced similar aliasing behavior for short targets on all the AMD CPUs tested (including Zen 3).

\subsection{Exploiting Unprivileged eBPF}

After BHI was disclosed to the Linux community, the upstream Linux kernel was updated to inline the indirect branch ``thunk'' calls where possible, including in JITed eBPF code. At the time of our research, this meant that the default configuration for AMD processors used \lfencejmp to protect indirect branches in unprivileged eBPF. To confirm whether the speculation window opened by \lfencejmp can be used in a realistic attack, we wrote a proof-of-concept (PoC) exploit which demonstrates that kernel memory contents can be inferred using eBPF on an AMD Zen 2 processor. Note that unprivileged eBPF has been disabled by default in recent kernels, as a consequence of the BHI disclosure~\cite{bhi}, therefore reproducing this PoC on current Linux kernels is only possible where unprivileged eBPF has been explicitly enabled.

The eBPF JIT translates code in a fairly direct manner from eBPF bytecode to x86 assembly, which gives us a certain amount of predictable control over register contents and memory contents. We obtain the needed indirect branch using the ``tail call'' mechanism, which transfers execution to another eBPF program. Before this indirect branch, a series of branches are executed in eBPF code so that the BHB at the indirect branch is constant, which is used by the indirect predictor to predict the target of the indirect branch. The target of the kernel branch is trained by executing an identical set of branches in a userspace application, with their addresses aliased to the corresponding kernel branches, as described above. We considered obtaining kernel addresses to be out-of-scope (recent academic work \cite{amdprefetch} has shown that fine-grained KASLR can be bypassed on AMD CPUs, and presumably cache side channels could be used to do similar on Intel CPUs), and instead used a small setuid program which printed the kernel address of our eBPF program.

We determined that, at the point of this indirect jump, the JITed code can store any value accessible to our eBPF program (including a kernel pointer) in R8 (eBPF Register 5), and an arbitrary user-controlled value in R11. For our PoC, we store a pointer to an eBPF map in R8, which is used as an \flushreload area for our cache-timing side channel, and store the kernel pointer to transiently read from in R11.

Although the aliasing on AMD CPUs allows a wide range of kernel code bytes to be used as targets, we did not search for suitable bytes that could be used as disclosure gadgets in existing kernel code. Instead, we created our own disclosure gadgets by embedding constant values inside eBPF code which would be interpreted and executed as x86 code when reached directly from a jump. Although Linux's eBPF code supports constant blinding to mitigate this, it is an optional hardening feature and is disabled by default. 

We used these constants to construct ``universal read'' gadgets which would infer a single bit of information from an arbitrary kernel address, and access a cache line based on whether that bit is 0 or 1. By creating one such gadget for every bit, we successfully used this PoC to infer the contents of kernel memory despite the use of the \lfencejmp mitigation, on Zen 2 with a suitable SMT workload.
Since the technique described above can be used by an attacker to construct arbitrary disclosure gadgets, our experiments imply that exploitation may also be possible without SMT.

We expect that the proof-of-concept BHI attacks \cite{bhi}, which use unprivileged eBPF, would also not be mitigated by \lfencejmp on Intel CPUs despite the use of eIBRS; these specific attacks seem likely to require an SMT workload due to the need for a larger speculation window (and 3 dependent loads), but this may not be the case for other BHI attacks.

\section{Related Work}

Our research builds upon the original work on Spectre~\cite{spectre}, including the detailed Google Project Zero writeup~\cite{spectregpz}, as well as research on newer variants of BTI-style attacks such as
SpectreRSB~\cite{spectrersb} (aka ret2spec~\cite{ret2spec})
and BHI~\cite{bhi}.
The possibility of a speculation window despite use of serialization, as in \lfencejmp, was suggested by Paul Turner as part of the motivation for retpoline~\cite{googleretpoline}.
Other mitigations for such attacks include
randpoline~\cite{randpoline} as well as a variety of more
targeted software and hardware defenses~\cite{evolutiondefenses}.

Research analyzing how BTI-style attacks and other building blocks can be used in practice also provides an improved understanding
of how hardware works, what software expects, and what protection
mitigations may need to provide. In particular,
SGXpectre~\cite{sgxpectre} investigated BTI attacks against SGX enclaves,
Zhang et al.~\cite{speculativetrojans} provided a detailed
analysis of branch predictors and other related microarchitecture
details of Intel CPUs,
and Mcilroy et al.~\cite{spectreisheretostay} analyze
transient execution vulnerabilities and 
argue that, at least for Chrome, process isolation is the
only realistic mitigation.
%Kirzner et al.\cite{speculativetypeconfusion}

There has also been a lot of recent research focusing on microarchitectural
side channels, and how they could be used in disclosure gadgets.
As well as BranchScope~\cite{branchscope}
and the LRU work by Xiong et al.~\cite{lrustates}, which we discussed in Section~\ref{sec:fittinginthewindow},
other key recent research includes
SMoTherSpectre~\cite{smotherspectre} which uses SMT port
contention as a side-channel rather than as an attack,
and NetSpectre~\cite{netspectre}, which uses timing
differences due to use of AVX2 instructions.
For a fairly comprehensive summary, we refer to
the survey of transient execution attacks and defenses
by Canella et al.~\cite{systematic}.
%TODO: Two methods for exploiting speculative control flow hijacks

\section{Conclusion}

We have shown that \lfencejmp is insufficient to mitigate BTI-style attacks on the tested Intel and AMD processors when SMT is enabled, in the absence of other mitigations. On the other hand, where SMT is not enabled (or not available), despite the presence of a speculation window, we were only able to transiently execute relatively short disclosure gadgets.
% Having said that, the absence of evidence is not evidence of absence.
Although one Intel processor as well as AMD processors prior to Zen 3 appear to have a speculation window large enough for two dependent loads without SMT, it remains an open question whether alternative disclosure gadgets may fit in this window on these or other processors.

In practice, the exploitability of BTI-style attacks depends on a range of factors beyond the existence of a speculation window, such as the availability of suitable indirect call sites and disclosure gadgets. However, where such attacks are a real concern, \lfencejmp alone does not seem to be sufficient to mitigate them, and other mitigations should be considered.

\section{Acknowledgements}

This work was partially inspired by people involved in the original Spectre response. In particular we would like to thank all the involved people at Google and Intel, many current and previous members of Intel STORM (in particular, Rodrigo Branco, Emma Benoit, Igor Chervatyuk, Lisa Aichele, and Thais Moreira Hamasaki), the Vrije Universiteit Amsterdam for their thought-provoking BHI research, the partners who motivated us to pursue this investigation, and finally AMD for their swift and friendly response to our findings involving their processors.

\section{Disclosure}

% This has the 'specifically focused on SMT' text that AMD wanted.
We reported our original findings (focused on SMT) involving AMD processors to AMD in November 2021 and followed principles of coordinated vulnerability disclosure with AMD. They agreed to align public disclosure with VU Amsterdam's BHI research (on March 8th, 2022), and assigned CVE-2021-26401 for the effectiveness issues on AMD parts. A previous version of this paper, specifically focused on SMT, was sent to AMD on February 23rd, 2022, and they independently reported the two-dependent-load case (without SMT) on AMD processors to the Linux community on February 25th, 2022; we have expanded our work to also include results for this case, in Section~\ref{sec:fittinginthewindow}.

%\section*{Bibliography}

\bibliographystyle{plain}
\bibliography{lfencejmp}

%%%%%%%%%%%%%%%%%%%%%%%%%%%%%%%%%%%%%%%%%%%%%%%%%%%%%%%%%%%%%%%%%%%%%%%%%%%%%%%%
\end{document}